\documentstyle[12pt]{article}
\def\tr#1{\,{\rm tr}\,#1\,}

\def\eop{\vspace*{\fill}\pagebreak}

\title{{\bf Loop Equations and Virasoro Constraints \\ in Matrix Models}
\vspace{2cm}}
\author{ Yu. Makeenko\thanks{E--mail: \ makeenko@itep.msk.su /
MAKEENKO@DESYVAX.bitnet \ or \ VXDESY::MAKEENKO}
\vspace{1cm}\\
{\it Institute for Theoretical and Experimental Physics}\\
{\it SU-117259 Moscow, USSR} }
\date{September, 1991}

\begin{document}

\maketitle

\begin{abstract}
In the first part of the talk, I review the applications of loop equations to
the matrix models and to 2-dimensional quantum gravity which is defined as
their
continuum limit. The results concerning multi-loop correlators for low genera
and the Virasoro invariance are discussed.

The second part is devoted to the Kontsevich matrix model which is equivalent
to
2-dimensional topological gravity. I review the Schwinger--Dyson equations for
the Kontsevich model as well as their explicit solution in genus zero.
The relation between the Kontsevich model and the continuum limit of
the hermitean one-matrix model is discussed.
\end{abstract}

\vspace{2cm}
\noindent
{\sl Talk at XXVth International Symposium on the
Theory of Elementary Particles, G\"osen, Germany, September 23--26}

\eop

\section{Introduction}

The relevance of matrix models to the problem of genus expansion of Feynman
graphs goes back to the original work by 't Hooft \cite{Hoo74}.
An explicit solution for the simplest case of the hermitean one-matrix model
had been first obtained by Br\'ezin, Itzykson, Parisi and Zuber \cite{BIPZ78}
in genus zero and then extended in \cite{Bes79,IZ80,BIZ80} to next few
genera.

The modern interest in matrix models is associated with
the context of statistical theories on random lattices and
discretized random surfaces \cite{Kaz85,Dav85,ADF85,KKM85} as well as with
the conformal field theory approach to 2D quantum gravity
\cite{Pol87,KPZ88,Dav88,DK89}. A connection between continuum limits of the
matrix model and minimal conformal models had been conjectured by Kazakov
\cite{Kaz89b} on the basis of genus-zero results.

The whole genus expansion of 2D quantum gravity has been constructed
in \cite{BK90,DS90,GM90a} taking the `double scaling limit' of
the (hermitean) one-matrix model. Moreover, the specific heat turns
out to obey a (non-perturbative) equation of the {\mbox Korteweg--de Vries}
type so that \sloppy
a relation between the continuum limit of the matrix models and integrable
theories emerges \cite{GM90b,BDSS90}.

While these results were obtained using orthogonal polynomial technique,
one more method --- that of loop (or Schwinger--Dyson) equations ---
is custom in studies of matrix models.
Loop equations had been proposed originally for
Yang-Mills theory both on a lattice \cite{Foe79,Egu79,Wei79} and in the
continuum \cite{MM79,Pol80} (for a review, see \cite{Mig83}) and then were
applied \cite{PR80,Fri81,Wad81} to matrix models.
A modern approach to loop equation which is based on its interpretation
as a Laplace equation on the loop space can be found in
\cite{Mak88,Mak89,HM89}.
The recent applications of loop equations to 2D quantum gravity have been
initiated by Kazakov \cite{Kaz89b}. The role of loops in 2D quantum gravity is
played by boundaries of a 2-dimensional surface.

Ref.\cite{Kaz89b} deals with genus zero. The whole set of loop equations
for 2D quantum gravity was first obtained by David \cite{Dav90} taking the
`double scaling limit' of the corresponding equations for the hermitean matrix
model. As was shown in \cite{Mig83,Dav90,AM90}, these equations can be
unambiguously solved order by order of genus expansion. However, this solution
is non-perturbatively unstable \cite{Dav90} as it should be for
2D euclidean quantum gravity.

One of the most interesting results which are obtained with the aid of loop
equations is the fact that the partition function of 2D quantum gravity
in an external background is the $\tau$-function of KdV hierarchy which is
subject to additional Virasoro constraints \cite{FKN91,DVV91a}. This proves a
conjecture of Douglas \cite{Dou90}.
The existence of Virasoro algebra was extended to the case of the matrix model
at finite $N$ in \cite{AJM90,GMMMO91,IM91a,MM90} while the relation to
the continuum Virasoro algebra of \cite{FKN91,DVV91a} had been studied by
${\rm  M}^4$ \cite{MMMM91}.

The second application of loop equations concerns the relation between 2D
quantum and topological \cite{LPW88,MS89} gravities. As Witten \cite{Wit90}
had conjectured, these two gravities are equivalent. This conjecture has been
verified in genus zero and genus one \cite{Wit90,DW90,Dis90} and
proven in \cite{FKN91,DVV91a} by showing that loop equations for 2D
quantum gravity coincide with the recursion relations between correlators  in
2D
topological gravity which were obtained by Verlindes \cite{VV91}.

 From the mathematical point of view, a solution of 2D topological gravity is
equivalent \cite{Wit90} to calculations of intersection indices on the moduli
space, ${\cal M}_{g,s}$, of curves of genus $g$ with $s$ punctures. Interesting
results for this problem have been obtained recently by Kontsevich \cite{Kon91}
who has represented the partition function of 2D topological gravity as that of
a (hermitean) matrix model in an external field. It is worth noting that the
Kontsevich matrix model is associated with the {\it continuum}\/ theory.
Therefore, it should be directly related to the `double scaling limit' of the
standard one-matrix model \cite{BK90,DS90,GM90a}.

The Kontsevich model can be studied by the method of loop equations.
As has been shown recently by Semenoff and the author \cite{MS91}, the
Schwinger-Dyson equations for the hermitean one-matrix model in an external
field, which is equivalent to the Kontsevich model, can be represented as a
set of Virasoro constraints imposed on the partition function. The large-$N$
solution of these equations, which is known from the work of Kazakov and
Kostov \cite{KK89}, solves the Kontsevich model in genus zero \cite{MS91}
showing explicitly the equivalence of 2D topological and quantum gravities to
this order.

The fact that the partition function of the Kontsevich model obeys the same
set of Virasoro constraints \cite{FKN91,DVV91a} as the continuum limit of the
hermitean one-matrix model has been proven recently by Witten \cite{Wit91}
using diagrammatic expansion and by (A.M.)$\mbox{}^3$ \cite{MMM91}
using the Schwinger--Dyson equations. This demonstrates an equivalence of 2D
topological and quantum gravities to any order of genus expansion.

In the first part of the talk, I review some works
\cite{AM90,AJM90,Mak90,MMMM91} on applications of loop equations
both to $N\times N$ matrix models at finite $N$ and to 2D quantum
gravity which is defined as their continuum limit.
The results for multi-loop correlators in low genera and
the Virasoro invariance both at finite $N$ and in the continuum are discussed.
The second part is devoted to the Kontsevich matrix model which is equivalent
to
2-dimensional topological gravity. The Schwinger--Dyson equations for
the Kontsevich model as well as their explicit solution in genus zero
\cite{MS91} is reviewed. Some original results concerning the Kontsevich
matrix model are reported.

\eop

\section{Matrix Models and 2D Quantum Gravity}

\subsection{Loop equation for hermitean matrix model}

The hermitean matrix model is defined by the partition function
\begin{equation}
Z^{\bf H}_N = \int   {\cal D}M\ \exp { -\tr{V(M)}}             \label{2.1}
\end{equation}
where $M$  is the $N\times N$  hermitean matrix. $V$  stands for a generic
potential
\begin{equation}
V(p) = \sum ^\infty _{k=0}t_kp^k .                        \label{2.2}
\end{equation}
The coupling  $t_k$ plays here the role of a source for the operator
$\tr{M^k}$
while  $V(p)$  is a source for the Laplace image of the Wilson loop
$\tr{[1/(p-M)]} $ :
\begin{equation}
\tr{V(M)} = \int ^{+i\infty +0}_{-i\infty +0}{dp\over 2\pi i} V(p) \,
\tr{{1\over p-M}} .                                          \label{2.3}
\end{equation}

The correlators  $\langle \tr{M^{k_1}}\ldots\tr{M^{k_m}}\rangle _c$ , where the
average is defined
with the same measure as in (\ref{2.1}), can be obtained differentiating
$\log{Z^{\bf H}_N}$ w.r.t.  $t_{k_1},\ldots,t_{k_m}$ while loop correlators can
be obtained by applying%
\footnote{\ According to this definition  $\delta V(p)/\delta V(q)=1/(p-q)$
which
plays the role of a $\delta $-function when integrated along imaginary axis.}
\begin{equation}
{\delta \over \delta V(p)} = -\sum ^\infty _{k=0}p^{-k-1}
{\partial \over \partial t_k}                             \label{2.4}
\end{equation}
so that the $m$-loop correlator reads
\begin{equation}
W^{\bf H}(p_1,\ldots,p_m) \equiv
\langle \tr{{1\over p_1-M}} \ldots \tr{{1\over p_m-M}}\rangle _c =
{\delta \over \delta V(p_1)} \ldots {\delta \over \delta V(p_m)}
\log{ Z^{\bf H}_N}.                                           \label{2.8}
\end{equation}
To calculate the actual values for the given model, say for the matrix model
with cubic potential, one should, after differentiations, put $t_k$'s equal
their actual values, say  $t_k=0$  for  $k>3$  in the case of cubic potential.

The loop equation can be derived using the invariance of the integral under an
(infinitesimal) shift of  $M$  and reads
\begin{equation}
\int _{C_1}{d\omega \over 2\pi i} {V'(\omega )\over (p-\omega) }
W^{\bf H}(\omega ) = (W^{\bf H}(p))^2 + {\delta \over \delta V(p)}
W^{\bf H}(p) .                                            \label{2.6}
\end{equation}
The contour $C_1$ encircles singularities of $W^{\bf H}(\omega )$ so that the
integration is a projector picking up negative powers of $p$. Eq.(2.6) is
supplemented with the asymptotic condition
\begin{equation}
pW^{\bf H}(p) \rightarrow  N\hbox{\ \  as }\ p \rightarrow  \infty
 \label{2.7}
\end{equation}
which is a consequence of the definition (\ref{2.8}).

Notice that one obtains the {\it single\/} (functional) equation for
$W^{\bf H}(p)$ . This is due to the fact that  $\tr{V(M)}$  contains a complete
set of operators. Such an approach is advocated in \cite{DVV91a,Mak90}. The set
of equations for multi-loop correlators (\ref{2.8}),
which is considered in \cite{Dav90,FKN91}, can be obtained from Eq.(\ref{2.6})
by $m-1$ --fold application of $\delta /\delta V(p_i)$.
The system of the standard Schwinger-Dyson
equations for the connected correlators  $\langle \tr{M ^{k_1}}\ldots \tr{M
^{k_m}}\rangle _c$ can be then obtained by expanding in powers of
$p^{-1}_1,\ldots,p^{-1}_m$ .

\subsection{Solution in $1/N$}

Eq.(\ref{2.6}) can be solved order by order of the expansion in $1/N^2$
(the genus
expansion). The second term on the r.h.s. represents the connected correlator
of two Wilson loops and is, in our normalization, of order 1 as
$N\rightarrow \infty $ while two other terms are of order  $N^2$ since
$W^{\bf H}(p)$  and  $V(p)$  are of order  $N$ . Therefore one can omit it as
$N\rightarrow \infty $  (which corresponds to genus zero = the planar limit).

The simplest (one-cut) solution of Eq.(\ref{2.6})  as $N\rightarrow \infty $
reads \cite{Mig83}
\begin{equation}
2W^{{\bf H}(0)}(p) = V'(p) - M(p)\sqrt{(\omega -x)(\omega -y)}    \label{3.3}
\end{equation}
where
\begin{equation}
M(p) = \int _{C_1}{d\omega \over 2\pi i}
{V'(p)-V'(\omega )\over (p-\omega )\sqrt{(\omega -x)(\omega -y)}}
 \label{3.4}
\end{equation}
is a polynomial of degree $K-2$ if  $V(p)$  is that of degree K. The ends
of the cut, $x$ and $y$, are determined from the asymptotics~(\ref{2.7}):
\begin{equation}
0 = \int _{C_1}{d\omega \over 2\pi i}
{V'(\omega )\over \sqrt{(\omega -x)(\omega -y)}}\hbox{ ;\ \  }     2N =
\int _{C_1}{d\omega \over 2\pi i}
{\omega V'(\omega )\over \sqrt{(\omega -x)(\omega -y)}} \equiv
{\cal W}(x,y) .      \label{3.2}
\end{equation}
For the even potential $(V(-p)=V(p))$, the first of these equations yields
$y=-x=\sqrt{z}$ which simplifies formulas. This case is called the
{\it reduced\/} hermitean matrix model.

The multi-loop correlators in the planar (genus zero) limit can be obtained by
varying according to the r.h.s. of Eq.(\ref{2.8}). The 2-loop correlator reads
\cite{AJM90}
\begin{equation}
W^{{\bf H}(0)}(p,q) = {1\over 4(p-q)^2} \left\lbrace {2pq -
(p+q)(x+y) +2xy}\over \sqrt{(p-x)(p-y)}\sqrt{(q-x)(q-y)}\right\rbrace
\label{3.5}
\end{equation}
while an expression for the 3-loop correlator is given in \cite{AJM90}. Note
that the 2-loop correlator~(\ref{3.5}) depends on the potential, $V$, only via
$x$ and $y$ but
not explicitly. This is not the case for all other multi-loop correlators.

To calculate $1/N^2$ correction to (\ref{3.3}) one needs
$W^{{\bf H}(0)}(p,p)$  which enters the r.h.s. of Eq.(\ref{2.6}).
Eq.(\ref{3.5})
yields
\begin{equation}
W^{{\bf H}(0)}(p,p) = {(x-y)^2\over 16(p-x)^2(p-y)^2} .
\label{3.6}
\end{equation}
and one can now obtain  $W^{{\bf H}(1)}(p)$  by an iteration of Eq.(\ref{2.6}).
The result
\begin{equation}
W^{{\bf H}(1)}(p) = {1\over \sqrt{(p-x)(p-y)}} \int _{C_1}{d\omega \over
{2\pi i}} {1\over {(\omega -p)M(\omega )}}
{(x-y)^2\over {16(\omega -x)^2(\omega -y)^2}} \label{3.7}
\end{equation}
is unambiguous \cite{Mig83,Dav90} provided that one requires analyticity of
$W^{{\bf H}(1)}(p)$  at zeros of $M(p)$. This procedure of iterative solution
can be pursued order by order of $1/N$-expansion.

\subsection{Continuum loop equation}

The continuum limit of the reduced hermitean matrix model is reached as
$N\rightarrow \infty $ while $K-1$ conditions  ${\cal W}^{(n)}(z_c)=0$ $(
{\cal W}(z)\equiv {\cal W}(-\sqrt{z},\sqrt{z}))$ with $n=1,\ldots,K-1$ are
imposed on the couplings, $t_k$, in addition to
(\ref{3.2}) at $K^{th}$ multi-critical point. 2D quantum gravity corresponds
to $K=2$. The `double scaling limit' can be obtained if one expands around the
critical point:
\begin{equation}
p \rightarrow  \sqrt{z_c} + {a\pi \over 2\sqrt{z_c}}\hbox{ ;\ }   z \rightarrow
\sqrt{z_c} - {a\sqrt{\Lambda }\over 2\sqrt{z_c}}, \label{4.1}
\end{equation}
so that $\pi $ and $\Lambda $ play the role of continuum momentum and
cosmological constant, respectively. The dimensionful cutoff $ a$ should
depend on $N$ such that the string coupling constant  $G=N^{-2}a^{-2K-1}$ would
remain finite as  $N\rightarrow \infty$  \cite{BK90,DS90,GM90a}.

To obtain the continuum limit of loop correlators (\ref{2.8}), it is
convenient to introduce the even parts
\begin{equation}
W^{even}(p_1,\ldots,p_m) \equiv
{\delta \over \delta V^{even}(p_1)}\ldots{\delta \over \delta V^{even}(p_m)}
\log \ Z^{reduced}_N \label{4.2}
\end{equation}
where  $Z^{reduced}_N$ and  $V^{even}(p)$  means, respectively,
(\ref{2.1}) and (\ref{2.2}) with  $t_{2k+1}= 0$ :
\begin{equation}
V^{even}(p) =\sum ^\infty _{k=0}t_{2k}p^{2k}\hbox{ ,\ }
{\delta \over \delta V^{even}(p)} = -\sum ^\infty _{k=0}p^{-2k-1}
{\partial \over \partial t_{2k}} .  \label{4.3}
\end{equation}
$W^{even}(p_1,\ldots,p_m)$  differs from  $W^{\bf H}(p_1,\ldots,p_m)$  by
correlators of products of traces of odd powers of $ M$. Near the critical
point, one gets
\begin{equation}
W^{\bf H}(p_1,\ldots,p_m) \rightarrow  2^{m-1} W^{even}(p_1,\ldots,p_m) .
 \label{4.4}
\end{equation}
This formula can be proven analyzing loop equations or by a direct inspection
of
multi-loop correlators \cite{AJM90}.

The continuum loop correlators can be obtained by the multiplicative
renormalization \cite{Dav90,AM90,FKN91}
\begin{equation}
W^{\bf H}_{2N}(p_1,\ldots,p_m) \rightarrow  2^m a^{-m} G^{{1\over2}m-1}
W_{cont}(\pi _1,\ldots,\pi _m)\hbox{\ \  for }\ m\geq 3
\label{4.5}
\end{equation}
while additional subtractions of genus zero terms are needed for $m=1$ and
$m=2$ :
\begin{equation}
W^{\bf H}_{2N}(p) - {1\over2}V'(p) \rightarrow  {1\over a\sqrt{G}}
( 2W_{cont}(\pi ) - J'(\pi ) )\hbox{ , }
\label{4.6}
\end{equation}
\begin{equation}
W^{\bf H}_{2N}(p_1,p_2) \rightarrow  4a^{-2} W_{cont}(\pi _1,\pi _2) +
{1\over a^2(\sqrt{\pi }_1+\sqrt{\pi }_2)^2 \sqrt{\pi _1\pi _2}}. \label{4.7}
\end{equation}
For latter convenience,  $W^{\bf H}_{2N}(p_1,\ldots,p_m)$  on the l.h.s.'s of
these formulas is the multi-loop correlator for the $2N\times 2N$ reduced
hermitean matrix model.

$J(\pi )$ on the r.h.s. of Eq.(\ref{4.6}) plays the role of a
source for the continuum Wilson loop:
\begin{equation}
W_{cont}(\pi _1,\ldots,\pi _m) = G
{\delta \over \delta J(\pi _1)}\ldots{\delta \over \delta J(\pi _m)}
\log \ Z_{cont} ,  \label{4.8}
\end{equation}
\begin{equation}
J(\pi ) = \sum ^\infty _{n=0}T_n\pi ^{n+{1\over2}}\hbox{ , }
{\delta \over \delta J(\pi )} = -\sum ^\infty _{n=0}\pi ^{-n-3/2}
{\partial \over \partial T_n} \label{4.9}
\end{equation}
with $T_k$ being sources for operators with definite scale dimension.
Therefore, Eqs.(\ref{4.5}), (\ref{4.7}) can be
derived from Eq.(\ref{4.6}) by varying w.r.t. $J(\pi )$.

The continuum loop equation can be obtained from (\ref{2.6})
substituting (\ref{4.6}), (\ref{4.7}):
\begin{equation}
\int _{C_1}{d\Omega \over 2\pi i} {J'(\Omega )\over (\pi -\Omega) }
W_{cont}(\Omega ) = (W_{cont}(\pi ))^2 + G
{\delta W_{cont}(\pi )\over \delta J(\pi )} + {G\over 16\pi ^2} +
{T_0^2\over 16\pi } . \label{4.10}
\end{equation}
This equation describes what is called the `general massive model'. It
corresponds to arbitrary $J(\pi )$ and interpolates between different
multi-critical points. For $K^{th}$ multi-critical point, one puts, after
varying w.r.t. $J(\pi )$, $T_n=0$ except for $n=0$ and $n=K$.

\subsection{Genus expansion}

Eq.(\ref{4.10}) can be solved order by order in $G$ (genus expansion)
analogously to that of Sect.2.2 . If $J(\pi )$ is a polynomial $(T_n=0$
for $n>K)$, $K-1$ lower coefficients of the asymptotic expansion of
$W_{cont}(\pi )$  are not fixed while solving in $1/\pi $ and should be
determined by requiring the one-cut analytic structure in $\pi $.
The continuum analog of (\ref{3.3}) reads
\begin{equation}
2W^{(0)}_{cont}(\pi ) = \int _{C_1}{d\Omega \over 2\pi i}
{J'(\Omega )\over (\pi -\Omega) } {\sqrt{\pi +u}\over \sqrt{\Omega +u}}
\label{5.1}
\end{equation}
where $u$ versus $\{T\}$ is determined from the asymptotic behavior.

This asymptotic relation can be obtained comparing $1/\pi $ terms in
Eq.(\ref{4.10}). Denoting the derivative w.r.t.  $x=-T_0/4$  by
$D$ , it is convenient to represent this relation as
\begin{equation}
2x = \int _{C_1}{d\Omega \over 2\pi i} J'(\Omega ) DW_{cont}(\Omega ).
\label{5.2}
\end{equation}
For the ansatz (\ref{5.1}), one gets
\begin{equation}
DW^{(0)}_{cont}(\pi ) = {1\over \sqrt{\pi + u} }
 - {1\over \sqrt{\pi }}.  \label{5.3}
\end{equation}

Eq.(\ref{5.3}) can be extended to any genera using the
representation \cite{GM90b,BDSS90}
\begin{equation}
DW_{cont}(\pi ) = 2\langle x| \left( \pi  + u(x) - {1\over 4}GD^2\right) ^{-1}
|x\rangle
- \ {1\over \sqrt{\pi }} = 2 \sum ^\infty _{n=1} R_n[u] \equiv 2R(\pi)
\label{5.4}
\end{equation}
where the diagonal resolvent of Sturm-Liouville operator is expressed via
the Gelfand-Diki\v{\i} differential polynomials \cite{GD75}
\begin{equation}
R_n[u]=2^{-n-1}\left( {G\over 8}D^2-u-D-D^{-1}uD\right)^n\cdot 1.
\label{R_n[u]}
\end{equation}
Substituting the r.h.s. of Eq.(\ref{5.4}) into Eq.(\ref{5.2}), one obtains
the string equation \cite{GM90b,BDSS90}
\begin{equation}
x = \sum ^\infty _{n=1}(n+{1\over 2})T_nR_n[u]. \label{5.5}
\end{equation}

The fact that the ansatz (\ref{5.4}) does satisfy Eq.(\ref{4.10}) is shown in
\cite{DVV91a}.
To this aid, one applies the operator
\begin{equation}
\Delta = -{G\over 16}D^4 + (u+\pi)D^2 + {1\over2}(Du)D,       \label{Del}
\end{equation}
which annihilates $W_{cont}(\pi)$ given by Eq.(\ref{5.4}), to Eq.(\ref{4.10}).
The result vanishes provided $u$ satisfies Eq.(\ref{5.5}) and
\begin{equation}
- 2 {\delta \over \delta J(\pi )} u = D^2W_{cont}(\pi )
\label{5.8}
\end{equation}
whose expansion in $1/\pi $ reproduces the KdV hierarchy
$\partial u/\partial T_n=DR_{n+1}[u]$ .

Comparing (\ref{4.8}) and (\ref{5.8}), one concludes that
\begin{equation}
Z_{cont} = \exp \left\lbrace - {2\over G}
\int^x _0{dy (x-y) u(y)} + \Phi(T_1,T_2,\ldots)\right\rbrace  \label{5.7}
\end{equation}
where the perturbative solution of Eq.(\ref{5.5}), which satisfies $u(0)=0$, is
chosen and an integration `constant' $\Phi(T_1,T_2,\ldots)$ depends on all
$T$'s
except for $T_0$. At the given multi-critical point, this $\Phi$ is unessential
so that the r.h.s. of Eq.(\ref{5.7}) coincides with the continuum partition
function which was obtained in \cite{BK90,DS90,GM90a} using the method of
orthogonal polynomials.

The general procedure of solving Eq.(\ref{4.10}) order by order in $G$ can be
now formulated as follows. One should first solve Eq.(\ref{5.5})
to find $u$ versus $x$ and $\{T_n\}$ (this is perturbatively unambiguous).
Then Eq.(\ref{5.4}) determines $DW_{cont}(\pi)$ while $W_{cont}(\pi)$ itself
can
be can be obtained by integrating
\begin{equation}
W_{cont}(\pi)= 2 \int ^x dx R(\pi)
\label{W_cont}
\end{equation}
where the integration `constant' can be  expressed via $\Phi$ entering
Eq.(\ref{5.7}). This constant becomes unessential for $K^{th}$ multi-critical
point when $T_n=0$ except for $n=0$ and $n=K$ so that $u$ depends only on the
cosmological constant $\Lambda$:
\begin{equation}
\Lambda ^{K/2} = {x4^{K+1}(K!)^2\over (2K+1)(2K)!T_K}.
\end{equation}

\subsection{Multi-loop correlators in 2D quantum gravity}

A formula which is similar to Eq.(\ref{W_cont}) exists for the multi-loop
correlators:
\begin{equation}
W_{cont}(\pi_1,\ldots,\pi_m) = 2 \int ^x dx {\delta \over \delta J(\pi_1)}
\cdots {\delta \over \delta J(\pi_{m-1})} R(\pi_m),
\label{Wmulti}
\end{equation}
where the integration `constant' depends again on $T_1,T_2,\ldots$ .

Since $R(\pi)$ depends on $T$'s only implicitly via $u$, the following chain
rule can be used for calculations
\begin{equation}
{\delta\over\delta J(\pi)}=2\int_{-\infty}^{+\infty}dx
DR(\pi){\delta\over\delta u(x)}
\label{xi}
\end{equation}
with ${\delta/\delta u(x)}$ being the standard variational derivative.
The expansion of Eq.(\ref{xi}) in $1/\pi$ reproduces the standard (commuting)
KdV flows.
An advantage of Eq.(\ref{xi}) is that it allows to obtain results without
solving the string equation (\ref{5.5}). Therefore, to calculate the multi-loop
correlator for a given multi-critical point, one can substitutes the solution
of Eq.(\ref{5.5}) only for this multi-critical point and should not solve it
for
arbitrary $T$'s.

An alternative way of calculating correlators in 2D quantum gravity is to take
the continuum limit (\ref{4.1}) of formulas of Sect.2.2 with the
aid of the renormalization \mbox{(\ref{4.5})--(\ref{4.7})}.
For the case of pure gravity (the $K=2$ critical point), the explicit form of
$W_{cont}(\pi )$ is
known for genus zero \cite{Dav90} and genus one \cite{AM90}:
\begin{equation}
W^{(0+1)}_{cont}(\pi ) = - {5\over 4}T_2 \left[ (\pi  -
{1\over2}\sqrt{\Lambda})
\sqrt{\pi +\sqrt{\Lambda }} \right] _- - {G\over 90T_2 }{(\pi +
{5\over2}\sqrt{\Lambda })\over \Lambda (\pi +\sqrt{\Lambda })^{5/2}}
\label{6.1}
\end{equation}
where $[\ldots]_-$ means subtraction of the term which diverges as
$\pi \rightarrow \infty $ as well as that of order $O(\pi ^{-{1\over2}})$.

The multi-loop correlators are known \cite{AM90,AJM90} for genus zero:
\begin{equation}
W^{(0)}_{cont}(\pi _1,\pi _2) = {\left(  \sqrt{\pi_1 + \sqrt{\Lambda }} -
\sqrt{\pi_2 + \sqrt{
\Lambda }}
\right) ^2\over 4(\pi _1-\pi _2)^2 \sqrt{\pi _1+\sqrt{\Lambda }}
\sqrt{\pi _2+\sqrt{\Lambda }}} -
{1\over 4(\sqrt{\pi _1}+\sqrt{\pi _2})^2\sqrt{\pi _1\pi _2}},
\label{6.2}
\end{equation}
\begin{equation}
W^{(0)}_{cont}(\pi _1,\ldots,\pi _m) \propto
{\partial ^{m-3}\over \partial \Lambda ^{m-3}} \left(  {1\over \sqrt{\Lambda }}
\prod ^{_m}_{i=1} {1\over (\pi _i+\sqrt{\Lambda })^{3/2}}\right) \hbox{
for }\ m\geq 3 .
\label{6.3}
\end{equation}
As is mentioned above, the additional subtraction is needed only for $m\leq 2.$

Analogous formulas can be obtained for higher multi-critical points. For $K=3$,
Eq.(\ref{6.2}) remain unchanged while the analog of Eq.(\ref{6.1}) reads
\cite{AM90}
\begin{equation}
W^{(0+1)}_{cont}(\pi ) = - {7\over 4}T_3 \left[ (\pi ^2 -
{1\over 2}\pi \sqrt{\Lambda } + {3\over 8}\Lambda )\sqrt{\pi +\sqrt{\Lambda }}
\right] _- - {4G\over 315T_3}
{ (\pi + {7\over4}\sqrt{\Lambda}) \over
\Lambda^{3/2}(\pi+\sqrt{\Lambda})^{5/2}}
.  \label{6.4}
\end{equation}

The above expressions for multi-loop correlators agree with those obtained
recently \cite{GL91,MMS91,MSS91,MS91b} for the Liouville theory.

\subsection{Complex matrix model}

The {\it complex\/} matrix model is defined by the partition function
\begin{equation}
Z^{\bf C}_N = \int {\cal D}M {\cal D}M^{\dag} \exp { -\tr{
V^{even}(MM^{\dag})}}
\label{7.1}
\end{equation}
where the integral goes over $N\times N$ complex matrices and  $V^{even}$ is
given by (\ref{4.3}). As is seen from this formula, the complex
matrix model resembles the reduced hermitean one. However, it differs by
combinatorics as well as by the fact that averages of traces of odd powers of
$M$ do not appear. The model (\ref{7.1}) has been studied in \cite{Mor91,AMP91}
using the orthogonal polynomial technique and in \cite{Mak90,AJM90} using loop
equations.

The variables $t_{2k}$ in (\ref{7.1}) play the role of sources
for the operators  $\tr{(MM^{\dag} )^k}$ while  $V^{even}$ is a source for the
Wilson loop  $\tr{[p/(p^2-MM^{\dag} )]}$ (comp.(\ref{2.3})):
\begin{equation}
\tr{V^{even}(MM^{\dag} )}
=\int ^{+i\infty +0}_{-i\infty +0}{dp\over 2\pi i} V^{even}(p) \,
\tr{{p\over (p^2-MM^{\dag})}} . \label{7.2}
\end{equation}
The analog of (\ref{2.8}) reads
\begin{equation}
W^{\bf C}(p_1,\ldots,p_m) = {\delta \over \delta V^{even}(p_1)}
\ldots{\delta \over \delta V^{even}(p_m)}
\log \ Z^{\bf C}_N   \label{7.3}
\end{equation}
which leads to the following loop equation for the complex matrix model
\cite{Mak90}
\begin{equation}
\int _{C_1}{d\omega \over 4\pi i} {V'^{even}(\omega )\over (p-\omega) }
W^{\bf C}(\omega ) = (W^{\bf C}(p))^2 + {\delta \over \delta V^{even}(p)}
W^{\bf C}(p) .\label{7.4}
\end{equation}
This equation should be supplemented with the asymptotic condition same as
(\ref{2.7}).

Comparison of Eq.(\ref{7.4}) and loop equation for the reduced
hermitean model yields for genus zero:
\begin{equation}
W^{\bf C}_N(p_1,\ldots,p_m) = {1\over 2} W^{even}_{2N}(p_1,\ldots,p_m) ,
\label{7.5}
\end{equation}
where the correlators  $W^{even}$ for the reduced hermitean model are defined
is Sect.2.3 . Notice that the correlator for $N\times N$ complex matrix model
enters the l.h.s. while that for $2N\times 2N$ reduced hermitean one enters
the r.h.s.. This guarantees the asymptotic condition
(\ref{2.7}). The coefficient $1/2$ in
Eq.(\ref{7.5}) leads to the following relation between the
partition functions for genus zero:
\begin{equation}
Z^{\bf C}_N \propto  \sqrt{Z^{reduced}_{2N}} .
\label{7.6}
\end{equation}

Due to the relation (\ref{7.5}),  $4W^{\bf C}(p)$  to the leading
order in $1/N$ is given by the r.h.s. of Eq.(\ref{3.3}) with
$V$  replaced by  $V^{even}$ and  $y=-x=\sqrt{z}$ . The multi-loop correlator
to leading order in $1/N$ can be then calculated by varying according to
(\ref{7.3}). The analog of (\ref{3.5}), (\ref{3.6}) reads
\begin{equation}
W^{{\bf C}(0)}(p,q) =
{1\over 4(p^2-q^2)^2}\left\lbrace {2p^2q^2-zp^2-zq^2\over \sqrt{p^2-z}
\sqrt{q^2-z}} - 2pq\right\rbrace\hbox{ ,  } W^{{\bf C}(0)}(p,p) =
{z^2\over 16p^2(p^2-z)^2} . \label{7.7}
\end{equation}
Moreover, an explicit expression for arbitrary multi-loop correlators exists
\cite{AJM90} for the complex matrix model even far from the critical point
(comp.(\ref{6.3})):
\begin{equation}
W^{{\bf C}(0)}(p_1,\ldots,p_m) = \left( {1\over {\cal W}'(z)}
{\partial \over \partial z}\right) ^{m-3}{1\over 2z{\cal W}'(z)}
\prod ^{_m}_{i=1} {z\over 2(p^2_i-z)^{3/2}}\hbox{\ \  for }\ m\geq 3 .
\label{7.8}
\end{equation}

As is proven in \cite{Mor91,AMP91,Mak90,AJM90}, the complex and hermitean
matrix
models belong to the
same universality class in the `double scaling limit'. This implies, in
particular, that the continuum limit of all multi-loop correlators coincide
with those of Sects.2.4,2.5 . They do not coincide, generally speaking,
for higher genera far from the critical points. Eqs.(\ref{7.5}),
(\ref{7.6}) remain valid, however, to arbitrary order of genus
expansion near the critical points. Using (\ref{4.4}), one concludes that
\sloppy $W^{\bf C}_N(p_1,\ldots,p_m)$  has the same continuum limit as
$2^{-m}W^{\bf H}_{2N}(p_1,\ldots,p_m)$  (i.e. the factor $2^m$ on the r.h.s.
of Eq.(\ref{4.5}) disappears for the complex matrix model) and
Eq.(\ref{7.4}) reproduces (\ref{4.10}).

\subsection{Loop equations as Virasoro constraints}

The loop equation (\ref{2.6}) can be represented as a set of Virasoro
constraints imposed on the partition function. Eq.(\ref{2.6}) can be rewritten,
using the definitions (\ref{2.2}) and (\ref{2.4}), as
\begin{equation}
{1\over Z^{\bf H}_N} \sum ^\infty _{n=-1} {1\over p^{n+2}} L^{\bf H}_n
Z^{\bf H}_N = 0
\label{8.1}
\end{equation}
where the operators\footnote{The dependence of these operators on $N$ is fixed
by $\partial \log Z_N/\partial t_0 = -N$.}\
\begin{equation}
L^{\bf H}_n = \sum ^\infty _{k=0}kt_k{\partial \over \partial t_{k+n}}
+  \sum _{0\leq k\leq n}
{\partial^2 \over \partial t_k \partial t_{n-k}}                  \label{8.2}
\end{equation}
satisfy \cite{AJM90,GMMMO91} Virasoro algebra
\begin{equation}
[L^{\bf H}_n,L^{\bf H}_m] = (n-m)L^{\bf H}_{n+m} .
\end{equation}
Therefore, Eq.(\ref{2.6}) is represented as the Virasoro constraints
\begin{equation}
L^{\bf H}_n Z^{\bf H}_N = 0\hbox{\ \      for }\ n\geq -1 .
\label{8.3}
\end{equation}
These constraints manifest the invariance of the integral on the r.h.s. of
(\ref{2.1}) under the shift of integration variable  $\delta M =
\epsilon \cdot M^{n+1}$ with  $n\geq -1$ \cite{AJM90,MM90}.

It is impossible, however, to make in (\ref{8.2}), (\ref{8.3}) the reduction to
even times. For even $n$, this reduction can be done for the first term on the
r.h.s. of (\ref{8.2}) but not for the second one. Therefore, there exist no
Virasoro constraints imposed on  $Z^{reduced}_N$ at \mbox{finite $N$}.

A set of Virasoro operators built up from the even times, $t_{2k}$,
arises for the complex matrix model. The loop equation (\ref{7.4}) can be
represented as Virasoro constraints
\begin{equation}
L^{\bf C}_n Z^{\bf C}_N = 0\hbox{\ \   for }\ n\geq 0;
\label{8.4}
\end{equation}
\begin{equation}
L^{\bf C}_n = \sum _{k=0}kt_{2k}{\partial \over \partial t_{2(k+n)}} +
\sum _{0\leq k\leq {n} }
{\partial^2 \over\partial t_{2k}\partial t_{2(n-k)}} .
\label{8.5}
\end{equation}
The Virasoro invariance is now related \cite{AJM90,MM90} to the change
$\delta M = \epsilon (MM^{\dag} )^{n}M$  with $n\geq 0$.

Analogously, the continuum loop equation (\ref{4.10}) can be represented as
Virasoro constraints which are imposed on  $Z_{cont}$ defined by (\ref{4.8}).
Using (\ref{4.9}), one proves that Eq.(\ref{4.10}) is equivalent to the
continuum Virasoro constraints \cite{FKN91,DVV91a}
\begin{equation}
{\cal L}^{cont}_n Z_{cont} = 0\hbox{\ \    for  }\ n\geq {-1} ;
\label{8.6}
\end{equation}
\begin{equation}
{\cal L} ^{cont}_n = \sum ^\infty _{k=0}(k+1/2)T_k
{\partial \over \partial T_{k+n}} + G \sum _{0\leq k\leq n-1}
{\partial ^2\over \partial T_k\partial T_{n-k-1}} + {\delta _{0,n}\over 16} +
{\delta _{-1,n}T^2_0\over 16G}. \label{8.7}
\end{equation}

The relation between the continuum Virasoro constraints (\ref{8.6}),
(\ref{8.7})
and those at finite $N$ can be studied \cite{MMMM91} without referring to loop
equations. Introducing
\begin{equation}
T_n = \sum _{k\geq n} {\sqrt{G}a^{n+1/2} kt_{2k}\Gamma (k+1/2)\over (k-n)!
\Gamma (n+3/2)} - 4N\sqrt{Ga} \delta _{n,0} ,
\label{8.8}
\end{equation}
or, vice versa,
\begin{equation}
kt_{2k} - 2N\delta _{k,0} = \sum _{n\geq k} (-)^{k-n} {a^{-n-1/2}
T_n\Gamma (n+3/2)\over \sqrt{G} (n-k)! \Gamma (k+1/2)} ,
\label{8.9}
\end{equation}
and rescaling the partition function
\begin{equation}
Z^{\bf C}_N \rightarrow  \tilde Z^{\bf C}_N = {\rm e} ^{- {1\over 2}
\sum \ A_{mn}\tilde T_m\tilde T_n} Z^{\bf C}_N ,
\label{8.10}
\end{equation}
\begin{equation}
A_{mn} = (-)^{n+m} {\Gamma (n+3/2)\Gamma (m+3/2)\over 2\pi (n+m+1)(n+m+2)n!m!}
G^{-1}a^{-m-n-1} ,
\label{8.11}
\end{equation}
one gets from (\ref{8.3}), (\ref{8.4})
\begin{equation}
\tilde {\cal L}^{\bf C}_n \tilde Z^{\bf C}_N = 0\hbox{\ \   for  } n=-1 ;\ \
\tilde {\cal L}^{\bf C}_n \tilde Z^{\bf C}_N = (-)^n{1\over 16a^n}
\tilde Z^{\bf C}_N\hbox{\ \  for }\ n\geq 0 ,
  \label{8.12}
\end{equation}
\begin{equation}
\tilde {\cal L}^{\bf C}_n  =
\sum ^\infty _{k=0}(k+1/2)T_k{\partial \over\partial \tilde T_{k+n}}+ G
\sum _{0\leq k\leq n-1}{\partial ^2 \over\partial \tilde T_k\partial
\tilde T_{n-k-1}}+{\delta _{n,0} \over16}+ {\delta _{n,-1}T^2_0 \over16G} .
\label{8.12pr}
\end{equation}
The variables \{$\tilde T$\} are related to \{$T$\} by
\begin{equation}
T_n = \tilde T_n + a {n\over n+1/2} \tilde T_{n-1} - 4N\sqrt{Ga} \delta _{n,0}
\label{8.13}
\end{equation}
so that the difference disappears as $a\rightarrow 0$.

Eqs.(\ref{8.8}), (\ref{8.9}) are the standard transition \cite{Kaz89b} to
operators with definite scale dimensions in the continuum. The rescaling
(\ref{8.10}) makes  $\tilde Z^{\bf C}_N$ finite as $a\rightarrow 0$ so that
$\tilde Z^{\bf C}_N \rightarrow  Z_{cont}$. While the operators  $\tilde {\cal
L}^{\bf C}_n$ tend to  ${\cal L} ^{cont}_n$ defined by (\ref{8.7}) as
$a\rightarrow 0$, the $a\rightarrow 0$ limit is not permutable with
differentiating  $\tilde Z^{\bf C}_N $ w.r.t. $T_n$ . This is why
$\tilde {\cal L}^{\bf C}_n \tilde Z^{\bf C}_N$ are nonvanishing (even
singular for $n\geq 1$) as $a\rightarrow 0$. These terms do not appear
\cite{MMMM91}, however, when  $\tilde {\cal L}^{\bf C}_n$ 's act on
\begin{equation}
e ^{- {1\over 2} \sum \ A_{mn}\tilde T_m\tilde T_n} \sqrt{Z^{reduced}_{2N}}
\rightarrow  Z_{cont}
\label{8.14}
\end{equation}
(comp. (\ref{7.6}), (\ref{8.10})). Thus, the l.h.s. of Eq.(\ref{8.14}) defines
the proper continuum partition function which is annihilated by the continuum
Virasoro operators (\ref{8.7}).

\eop

\section{Kontsevich Model and 2D Topological Gravity}

\setcounter{equation}{0}
\subsection{2D topological gravity as Kontsevich model}

The starting point in demonstrating an equivalence between 2D topological
gravity and the Kontsevich model is the Witten's geometric formulation
\cite{Wit90} of 2D
topological gravity. In this formulation, one calculates the correlation
functions of $s$ operators \sloppy
$\sigma_{n_1}(x_1),\ldots,\sigma_{n_s}(x_s)$
with definite (non-negative integer) scale dimensions $n_i$, living on
a 2-dimensional Riemann surface $\Sigma$ of genus $g$. Those are expressed
\cite{Wit90} via the intersection indices
\begin{equation}
\left\langle \sigma_{n_1}(x_1)\cdots\sigma_{n_s}(x_s) \right\rangle _g=
\int \prod_i c_1({\cal L}_{(i)})^{n_i} {\cal N}(n_i)
\label{corr}
\end{equation}
where $c_1({\cal L}_{(i)})$ is the first Chern class of the line bundle (which
is the cotangent space to a curve at $x_i$) over the moduli space, ${\cal
M}_{g,s}$, of curves of genus $g$ with $s$ punctures and the integral goes
over $\bar{\cal M}_{g,s}$. The normalization factor $\prod_i {\cal N}(n_i)$,
which is related to the normalization of the operators $\sigma$, is to be fixed
below. The r.h.s. of Eq.(\ref{corr}) is non-vanishing only if
\begin{equation}
\sum_i n_i = 3g-3+s,  \label{dimension}
\end{equation}
{\it i.e.} the (complex) dimension of ${\cal M}_{g,s}$. Notice the crucial
property of correlators in topological theories --- those depend only on the
dimensions $n_1,\ldots,n_s$ and genus $g$ but not on the metric on $\Sigma$
and, therefore, not on positions of the punctures $x_1,\ldots,x_s$.

It is convenient to introduce the set of couplings $t_n$ which play the role
of sources for the operators $\sigma_n$. The genus $g$ contribution to the free
energy then reads
\begin{equation}
F_g[t]= \left\langle \exp{\left( \sum_n t_n \int \sigma_n\right)}
\right\rangle_g
\label{F_g}
\end{equation}
while the correlator on the l.h.s. of Eq.(\ref{corr}) can be obtained by
differentiating $F_g[t]$ w.r.t. $t_{n_1},\ldots,t_{n_s}$ since the correlators
do not depend on $x_1,\ldots,x_s$. The total free energy can be obtained
from the genus expansion
\begin{equation}
F[t;\lambda]= \sum_g \lambda^{2g-2} F_g[t]
\label{F}
\end{equation}
with $\lambda^2$ being the string coupling constant%
\footnote{We use in this part of the talk the Witten's normalization
\cite{Wit90} of 2D topological gravity (${\cal N}(n)=n!$ in Eq.(\ref{corr})).
It is related to the normalization \cite{BDSS90} of the matrix models
(${\cal N}(n)=(2n+1)!!$), which is used in the first part, as follows:
$\lambda^2=2G,\ n!\,t_n=(2n+1)!!\,T_n+\delta_{1n}$}.
Note, that due to the
relation (\ref{dimension}), the $\lambda$-dependence of $F$ can be absorbed
by the rescaling of $t_n$:
\begin{equation}
F[\ldots,t_n,\ldots;\lambda] = F[\ldots,\lambda^{{2\over 3}(n-1)}
t_n,\ldots;1].
\label{rescale}
\end{equation}

The Kontsevich approach \cite{Kon91} to evaluate $F[t;\lambda]$, given by
Eq.(\ref{F}), is based on a combinatorial calculation of the intersection
indices on ${\cal M}_{g,s}$. Let us represent Eq.(\ref{F_g}) as
\begin{equation}
F_g[t]= \sum_{s\geq 0} \sum_{n_1,\ldots,n_s}
{1\over s!}F_{g,s}^{\,(n_1,\ldots,n_s)} \hbox{ }
\label{bandexp}
\end{equation}
where
\begin{equation}
F_{g,s}^{\,(n_1,\ldots,n_s)} =
\left\langle \sigma_{n_1}(x_1)\cdots\sigma_{n_s}(x_s) \right\rangle _g
t_{n_1}\cdots t_{n_s}.
\label{band}
\end{equation}
The last quantity can be interpreted as a contribution from a {\it band\/}
graph
(or a {\it fat}\/ graph in Penner's terminology \cite{Pen88})
of genus $g$ with $s$ loops and three bands linked at each vertex.
These graphs were introduced in quantum field theory by 't Hooft \cite{Hoo74}.
The original Riemann surface with $s$ punctures can be obtained from this band
graph by shrinking the boundaries of bands (forming loops) into the punctures.

As is proven by Kontsevich \cite{Kon91},
\begin{equation}
\sum_{n_1,\ldots,n_s}
\left\langle \sigma_{n_1}(x_1)\cdots\sigma_{n_s}(x_s) \right\rangle _g
\prod_i {(2n_i-1)!!\over {\cal N}(n_i)} \tr{\Lambda^{-2n_i-1}}
=\sum_{graphs_{g,s}}{2^{-\#(vert.)}\over \#(aut.)} \prod_{links_{i,j}}
{2\over \Lambda_i+\Lambda_j}
\label{cell}
\end{equation}
where $\Lambda_i$ are eigenvalues of a $N\times N$ hermitean matrix $\Lambda$
and the sum goes over the connected band graphs with \#(vert.) vertices. The
product goes over the links of the graph. Each link carries two indices $i,j$
which are continuous along the loops while each of $s$ traces on the l.h.s.
corresponds to the summation over the index along one of $s$ loops. The
combinatorial factor \#(aut.) in the denominator is due to a symmetry of the
graph.

Substituting Eq.(\ref{cell}) into Eqs.(\ref{band}),(\ref{bandexp}), identifying
\begin{equation}
t_n={\lambda (2n-1)!!\over {\cal N}(n)} \,\hbox{tr}\,(\Lambda^{-2n-1})
\label{ident}
\end{equation}
and making use
of Eq.(\ref{rescale}), one represents the r.h.s. of Eq.(\ref{F})
in the form of the logarithm of the partition function
\begin{equation}
Z_{Konts}[\Lambda;\lambda]
\equiv {\int{\cal D}X\hbox{e}^{\tr{({\sqrt{\lambda}\over 6}X^3 -
{1\over2}\Lambda X^2)}} \over
\int{\cal D}X\hbox{e}^{-{1\over2}\tr{\Lambda X^2}}}
\label{Zlambda}
\end{equation}
where the integral goes over the hermitean $N\times N$ matrix $X$. The original
normalization of \cite{Kon91} corresponds to $\lambda=-1$.

There is, however, a {\it subtlety} in the identification (\ref{ident}).
The point is that, for an $N\times N$ matrix
$\Lambda^{-1}$, $\tr{(\Lambda^{-k})}$ are independent  only for $1 \leq k \leq
N$ while, say $\tr{(\Lambda^{-N-1})}$ is reducible. All
$\tr{(\Lambda^{-2n-1})}$ become independent, as it should be for the sources in
2D topological gravity, as $N\rightarrow\infty$. Therefore
\begin{equation}
\log{Z_{Konts}[\Lambda;\lambda]}\rightarrow F[t; \lambda].
\label{logZ}
\end{equation}
only as $N\rightarrow\infty$.
The equality (\ref{logZ}) is valid in a sense of an asymptotic
expansion at large $\Lambda$ with each term being finite providing $\Lambda$
is positively defined.

Let us explain the Kontsevich results from the viewpoint of the standard
analysis of the matrix model (\ref{Zlambda}). $Z_{Konts}[\Lambda;\lambda]$
admits the perturbative expansion in $\lambda$ that starts from the term
$O(\lambda)$. This term corresponds to the contribution of three puncture
operators in genus zero \cite{Wit90}:
\begin{equation}
F[t;\lambda]= {t_0^3\over 6\lambda^2}+\ldots\ ,
\label{t3over6}
\end{equation}
where $\lambda^2$ in the denominator emerges because of Eq.(\ref{ident}).
The contribution of a generic graph with \#(vert.) vertices, \#(link) links
and $s$ loops is proportional to
\begin{equation}
\lambda^{\#(vert.)\over 2} N^s = (\lambda N)^s\lambda^{2g-2}
\end{equation}
in an agreement with Eq.(\ref{cell}).

Notice that while $N\rightarrow\infty$,
all terms of the perturbative expansion in $\lambda$
contribute to $F[t;\lambda]$ in
contrast to the standard large-$N$ expansion by 't Hooft \cite{Hoo74} when an
expansion in ${1\over N^2}$ emerges so that $N=\infty$ corresponds to planar
graphs only. The 't Hooft case can be reproduced if $\lambda \sim N^{-1}$.
Then $F\sim N^2$ while
\begin{equation}
W_{Konts}[\Lambda;{1\over N}]\equiv {1\over N^2}
\log{Z_{Konts}[\Lambda;{1\over N}]}\rightarrow{1\over N^2} F[t;{1\over N}]
\label{WKonts}
\end{equation}
is finite. Therefore,
$W_{Konts}[\Lambda;0]=F_0[t]$ can be obtained in the 't~Hooft planar limit.
This fact has been utilized in \cite{MS91} to solve the Kontsevich model
in genus zero.

\subsection{The Schwinger-Dyson equations}

The Kontsevich model can be studied using the custom methods of solving matrix
models. Since $\Lambda$ in Eq.(\ref{Zlambda}) is a matrix, the standard
orthogonal polynomial technique can not be applied.
For this reason, the method of Schwinger-Dyson equations has been applied to
this problem \cite{MS91,MMM91}.

To derive the Schwinger--Dyson equations, it is convenient to make a linear
shift of the integration variable in the numerator on the r.h.s. of
Eq.(\ref{Zlambda}). Modulo an unessential constant, one gets
\begin{equation}
Z_{Konts}[\Lambda;\lambda]=  \prod_i \sqrt{\Lambda_i} \prod_{i>j}
(\Lambda_i+\Lambda_j)\,\hbox{e}^{-{\Lambda^3\over 3\lambda}}
Z\left[{\Lambda^2\over (2\lambda)^{2\over 3}} \right]
\label{Gauss}
\end{equation}
where
\begin{equation}
Z[M]= \int {\cal D}X\ \hbox{e}^{\tr{(-{X^3\over 3}+MX)}}
\label{Z[M]}
\end{equation}
and the Gaussian integral in the denominator has been calculated.

$Z[M]$ which is defined by Eq.(\ref{Z[M]}) is the standard partition function
of the hermitean one-matrix model in an $N\times N$ matrix external field $M$.
This external field problem, which is analogous to the corresponding problem
\cite{BN81,BG80,BRT81} for the unitary matrix model, has been studied recently
in \cite{GN91,MS91}. While the representation (\ref{Zlambda}) is
convenient for constructing the perturbation theory expansion,
the partition function $Z[M]$ is convenient for deriving Schwinger-Dyson
equations.

The partition function (\ref{Z[M]}) depends on $N$ invariants, $m_i$, ---
the eigenvalues of $M$. Let us perform the integral over angular
variables in the standard way \cite{IZ80,Meh81}
to express $Z[M]$ as the integral over $x_i$ --- the eigenvalues
of $X$. Modulo an irrelevant multiplicative constant, the result reads
\begin{equation}
Z[m]=\int \prod _{i=1} ^N dx_i{\Delta[x] \over \Delta[m] } \exp{ \sum _{i=1} ^N
(-{x_i^3\over 3} + m_ix_i)}
\label{Z[m]}
\end{equation}
where $\Delta[m]= \prod _{i< j} (m_i-m_j)$ is the Vandermonde determinant.

The Schwinger--Dyson equations result from the following change of
variables in (\ref{Z[m]}): $x_i \rightarrow x_i+ \epsilon _n x_i^{n+1}$ for
$i=1, \ldots, N$ and $n\geq -1$ in full analogy to the matrix model without
external field \cite{AJM90,IM91a,MM90}. Noticing that $x_i$ in the integrand
can
be replaced
by ${\partial \over \partial m_i}$ when applied to $\Delta[m]Z[m]$, the set of
Schwinger--Dyson equations can be written in the form \cite{MS91}
\begin{equation}
L_n \Delta[m]Z[m] =0     \hbox{  \ \ for \ \ \ } n\geq -1
\label{constr}
\end{equation}
with
\begin{equation}
L_n= \sum _i \left\lbrace
- \left({\partial \over \partial m_i}\right)
^{n+3} + \left({\partial \over \partial m_i}\right) ^{n+1} m _i +
{1\over 2}\sum _{k=0} ^n \sum _{j\neq i} \left({\partial \over \partial m_i}
\right)^k
\left({\partial \over \partial m_j}\right)^{n-k} \right\rbrace.
\label{L_n}
\end{equation}
It is easy to verify by a direct calculation that these operators obey
Virasoro algebra
\begin{equation}
[L_n,L_m]=(n-m)L_{n+m}.
\label{comm}
\end{equation}

The Virasoro generators (\ref{L_n}) annihilate the totally antisymmetric
function $\Delta[m] Z[m]$. One can easily construct the generators ${\cal L}_n$
which annihilate $Z[m]$ itself. Let us introduce for this purpose the `long'
derivatives
\begin{equation}
\nabla _i \equiv \Delta^{-1}[m] {\partial \over \partial m_i} \Delta[m]
= {\partial\over \partial m_i} + \sum_{j \neq i}{1 \over m_i-m_j}
\label{nabla}
\end{equation}
which commute one with each other. The Virasoro constraints (\ref{constr}),
(\ref{L_n}) now take the form \cite{MS91}
\begin{equation}
{\cal L}_n Z[m] =0     \hbox{  \ \ for \ \ \ } n\geq -1
\label{calconstr}
\end{equation}
and
\begin{equation}
{\cal L}_n= \sum _i \left\lbrace -(\nabla _i)^{n+3} + (\nabla _i) ^{n+1} m _i +
{1 \over 2} \sum _{k=0} ^n \sum _{j \neq i} (\nabla _i)^k (\nabla _j)^{n-k}
\right\rbrace
\label{calL_n}.
\end{equation}
Due to the commutativity of $\nabla$'s, the order in the last term in
unessential. As follows from the definition (\ref{nabla}), the generators
(\ref{calL_n}) obey the Virasoro algebra commutation relations, same as
(\ref{comm}).

The Virasoro constraints (\ref{calconstr}), (\ref{calL_n}) turn out to be
equivalent to the following equation
\begin{equation}
\left\lbrace (\partial _i)^2 + \sum _{j \neq i} {1 \over
m_i-m_j}(\partial_i-\partial_j) - {m_i \over N} \right\rbrace Z[m] =0
\label{me}
\end{equation}
which is called in \cite{MS91} the `master equation'. As is shown in
\cite{GN91,MMM91}, Eq.(\ref{me}) results from shifting $X$ in
Eq.(\ref{Z[M]}) by an arbitrary (hermitean) matrix while Eqs.(\ref{constr}),
(\ref{L_n}) (or (\ref{calconstr}), (\ref{calL_n})) result from the
shift $X \rightarrow X + \epsilon _n X^{n+1}$.

\subsection{The genus-zero solution}

Eq.(\ref{me}) can be solved in genus zero using the standard methods of
the large-$N$ limit. As $N\rightarrow\infty$, Eq.(\ref{me}) is reduced to an
integral equations which is similar to those solved by Br\'ezin and Gross
\cite{BG80} with the aid of the Riemann--Hilbert method. The corresponding
solution had been first found by Kazakov and Kostov \cite{KK89} and is
discussed in \cite{GN91,MS91}.

Substituting this solution into Eq.(\ref{Gauss}) which expresses the partition
function of the Kontsevich model via that for the hermitean matrix in an
external field, one gets in genus zero
\begin{eqnarray}
F_0
&=& {1 \over N}\sum_i \left\lbrace {1 \over 3} (\Lambda_i^2-2u)^{3\over 2}
+u\sqrt{\Lambda_i^2-2u}+{u^3\over 6} - {\Lambda_i^3\over 3}\right. \nonumber \\
& & \mbox{}+\left.{1 \over 2N}\sum_j \left[ \log{\left(\Lambda_i+
\Lambda_j\right)}-
\log{\left(\sqrt{\Lambda_i^2-2u}+\sqrt{\Lambda_j^2-2u}\right)} \right]
 \right\rbrace
\label{KKform}
\end{eqnarray}
where $u[\Lambda]$ is determined by
\begin{equation}
u = {1\over N}\sum _i {1 \over \sqrt{(\Lambda_i^2-2u)}} .
\label{u}
\end{equation}

The solution (\ref{KKform}) is similar to the strong coupling solution of
\cite{BG80} while Eq.(\ref{u}) has emerged to guarantee correct analytic
properties. It is important that the r.h.s. of Eq.(\ref{KKform}) is stationary
w.r.t. $u$ due to Eq.(\ref{u}).
For a constant field $\Lambda _i = (6g)^{-{2\over 3}}$, this solution recovers
the results of Br\'ezin et al. \cite{BIPZ78} for the case of a cubic
interaction.

Eq.(\ref{u}) can be rewritten in the form of the string equation of a `general
massive model' \cite{BDSS90} in genus zero. To this aim, let us expand the
r.h.s. of Eq.(\ref{u}) in $u$ and substitute
\begin{equation}
t_n = {1\over N} {(2n-1)!!\over n!} \sum_i \Lambda_i^{-2n-1}.
\label{t2DTG}
\end{equation}
This equation is nothing but Eq.(\ref{ident}) with ${\cal  N}(n)=n!$ which
fixes the normalization \cite{Wit90} of 2D topological gravity
and $\lambda={1\over N}$ as is prescribed by Eq.(\ref{WKonts}).
We rewrite Eq.(\ref{u}) finally as
\begin{equation}
 u = \sum_{n=0}^{\infty} t_n u^n .
\label{se0}
\end{equation}
The precise form of the genus-zero string equation can be obtained by the
well-known shift~\cite{DW90}: $t_1\rightarrow t_1+1$.

\subsection{Relation to 2D topological and quantum gravities}

It is instructive to compare the solution (\ref{KKform}) of the Kontsevich
model with known results for the partition functions of 2D topological and
quantum gravities in genus zero.
To obtain the perturbative expansion of $F_0[t]$, one solves Eq.(\ref{se0})
by iterations in $u$:
\begin{equation}
u=t_0+t_0t_1+t_0t_1^2+t_0^2t_2 +\ldots
\label{upert}
\end{equation}
with $t_n\sim t_0^{2n+1}$, and substitutes the result into the r.h.s. of
Eq.(\ref{KKform}) which is expanded in $u$ with Eq.(\ref{t2DTG}) has been used.

While the complicated structure of the perturbative expansion of $F_0[t]$
represents the variety of planar band graphs (taken with appropriate
combinatorial coefficients), great simplifications occur for derivatives of
$F_0[t]$. Let us define ${\cal D}$ by
\begin{equation}
{\cal D}=2\sum_i {\partial\over \partial(\Lambda_i^2)}.
\label{D}
\end{equation}
Then, by a direct differentiation of Eq.(\ref{KKform}), one gets
\begin{equation}
{\cal D} F_0= {t_0^2\over 2}+{u^2\over 2}-\sum_{k=0}^{\infty}t_k
{u^{k+1}\over (k+1)}.
\label{DF0}
\end{equation}
This expression is again stationary w.r.t. $u$ due to Eq.(\ref{se0}) so that
one more application of ${\cal D}$ yields
\begin{equation}
{\cal D}^2 F_0=u-t_0(1+t_1).
\label{susc}
\end{equation}

To compare Eq.(\ref{KKform}) with the known solution of 2D topological gravity
in genus zero, let us notice that ${\cal D}$ defined by (\ref{D}) can be
rewritten using Eq.(\ref{t2DTG}) as
\begin{equation}
{\cal D}= - \sum_{n=1} ^\infty nt_n{\partial\over \partial t_{n-1}}.
\label{Dt}
\end{equation}
This is exactly the operator entering the puncture equation \cite{DW90}
which reads in genus zero:
\begin{equation}
{\partial F_0\over \partial t_0}= {t_0^2 \over 2} - {\cal D}F_0.
\label{pe}
\end{equation}
Therefore, one gets from Eq.(\ref{DF0})
\begin{equation}
{\partial F_0\over\partial t_0}= \sum_{k=0}^{\infty}t_k
{u^{k+1}\over (k+1)} -{u^2\over 2}
\label{dF0dt0}
\end{equation}
which is a true formula that gives in particular
\begin{equation}
{\partial^2 F_0\over \partial t_0^2} = u .
\label{defu}
\end{equation}
Since $t_0$ is the cosmological constant, one sees from this formula $u$ to be
the string susceptibility.

Using Eq.(\ref{defu}), one can immediately calculate the critical index
$\gamma_{string}$.
For $K^{th}$ multi-critical point, when one puts all $t_n=0$ except for $n=0$
and $n=K$, Eq.(\ref{se0}) yields $u \propto x^{1 \over K}$ ,
$\gamma_{string}=-{1\over K}$ in full analogy to \cite{Kaz89b}.
Notice, however, that the solution (\ref{KKform}) is associated with the {\it
continuum\/} interpolating model while in the standard case one `interpolates'
by a matrix model whose couplings should be turned to critical values in
order to reach the continuum limit.

Finally, let us mention that the genus-zero solution (\ref{KKform})
can be rewritten exactly in the form of that
for the hermitean one-matrix model in the continuum limit. Let us first note
that Eq.(\ref{dF0dt0}) can be viewed as an integrated version of
Eq.(\ref{defu}):
\begin{equation}
{\partial F_0\over \partial t_0}= \int _0^{t_0} dx u(x)
\label{integral0}
\end{equation}
where $u(t_0)$ is a solution of Eq.(\ref{se0}) which is considered as a
function
of $t_0$ at fixed values of $t_n$ for $n\geq 1$. The integration constant is
fixed by the fact that $u(0)=0$ which is nothing but the condition that
chooses the perturbative solution of the string equation.

One more integration of Eq.(\ref{integral0}) yields
\begin{equation}
F_0 = \int _0^{t_0} dx (t_0-x) u(x) + \Phi(t_1,t_2,\ldots) .
\label{tau0}
\end{equation}
which coincides with the
representation (\ref{5.7}) of the free energy for 2D quantum gravity.
As has been proven recently \cite{Wit91,MMM91}, the Kontsevich model obeys the
same set of Virasoro constraints (\ref{8.6}), (\ref{8.7}) as 2D quantum
gravity.
This demonstrates an equivalence of 2D topological and quantum gravities to
arbitrary genus.

\section{Concluding remarks}

Loop equations turned out to be a useful tool in studies of matrix models as
well as of their continuum limit associated with 2D quantum gravity with
matter.
The point is that loop equations are literally the
Virasoro constraint
imposed on the partition function. In the continuum limit, this Virasoro
symmetry represents the underlying conformal invariance.

The appearance of new symmetries of loop equations (as well as the very idea of
the `double scaling limit' \cite{BK90,DS90,GM90a}) is very interesting from the
viewpoint of multi-dimensional loop equations (see \cite{Mig83}).
A step along this line has been done in \cite{FKN91,DVV91a,Goe91} where
the $W$-algebras were associated with the continuum limit of multi-matrix
models. It would be interesting to find an analog of this symmetry for
multi-matrix models at finite $N$.

The fact that the Kontsevich matrix model is a solution of the continuum
Virasoro constraints (and, therefore, of the continuum loop equations) throws
light on the origin of Virasoro constraints. It is non-trivial that this matrix
model appears as a solution of the continuum loop equation. This seems to be
analogous to the known property \cite{Mig83} of multi-dimensional loop
equations which possess solutions differing from the original Yang--Mills path
integral.

\eop

\end{document}